\documentclass[12pt]{article}

\font\grande=cmr9.5 scaled \magstep4
\font\medio=cmr9.5 scaled \magstep2
\outer\def\beginsection#1\par{\medbreak\bigskip
      \message{#1}\leftline{\bf#1}\nobreak\medskip
\vskip-\parskip
      \noindent}
\usepackage{graphicx} 
\begin{document}
\bibliographystyle {unsrt}

\titlepage

\begin{flushright}
CERN-PH-TH/2009-028
\end{flushright}

\vspace{15mm}
\begin{center}
{\grande Estimating relic magnetic fields}\\
\vspace{5mm}
{\grande from CMB temperature correlations}\\
\vspace{1.5cm}
 Massimo Giovannini 
 \footnote{Electronic address: massimo.giovannini@cern.ch} \\
\vspace{1cm}
{{\sl Department of Physics, 
Theory Division, CERN, 1211 Geneva 23, Switzerland }}\\
\vspace{0.5cm}
{{\sl INFN, Section of Milan-Bicocca, 20126 Milan, Italy}}
\vspace*{2cm}

\end{center}

\vskip 1cm
\centerline{\medio  Abstract}
The  temperature and polarization inhomogeneities of the Cosmic Microwave Background  might bear the mark of pre-decoupling magnetism. The parameters of 
a putative magnetized background are hereby estimated from the observed temperature autocorrelation as well as from the measured temperature-polarization cross-correlation.  
\noindent

\vspace{5mm}

\vfill
\newpage
In recent years the observations of the Cosmic Microwave Background (CMB in what follows) have been translated into a set of constraints on the standard parameters of an underlying $\Lambda$CDM scenario where $\Lambda$ stands for the dark energy component (parametrized in terms of a cosmological constant) and CDM is the acronym of cold dark matter which is the dominant source of non-relativistic matter of the model. More specifically, 
the five year release of the WMAP data \cite{WMAP5} (WMAP 5yr in what follows) can be used to infer the parameters of a putative $\Lambda$CDM paradigm where $\Omega_{X\,0} = \rho_{X 0}/\rho_{\mathrm{crit}}$ denotes the (present) energy density of the species $X$ in units of the critical energy density $\rho_{\mathrm{crit}}$. The 
best-fit values of the $\Omega_{X0}$ as well as of the other parameters
 of the model can be determined, for instance, from the WMAP 5yr data alone  \footnote{The combination of the WMAP 5yr data with other cosmological data sets 
(e.g. supernovae \cite{SN} and large scale structure \cite{LSS} observations) lead to compatible values \cite{WMAP5}.} (see also \cite{WMAP3} for earlier results) and they are 
\begin{equation}
\Omega_{\mathrm{b}0} = 0.0441 \pm 0.0030,\qquad \Omega_{\mathrm{c}0} = 0.214 \pm 0.027,\qquad \Omega_{\Lambda0} =0.742\pm 0.030,
 \label{int1}
 \end{equation}
 where the subscripts b and c stand, respectively, for baryons and CDM components; Eq. (\ref{int1}) does not exhaust the list of parameters which also include the Hubble constant (i.e. $H_{0}$), the optical depth of reionization (i.e. $\epsilon_{\mathrm{re}}$) and  the spectral index of (adiabatic) curvature perturbations (i.e. $n_{\mathrm{s}}$):
\begin{equation}
 H_{0} = 71.9^{2.6}_{-2.7}\, \mathrm{km/sec} \, \mathrm{Mpc}^{-1},\qquad \epsilon_{\mathrm{re}} = 0.087 \pm 0.017, \qquad
 n_{\mathrm{s}} = 0.963_{-0.015}^{0.014};
\label{int2}
\end{equation}
$\epsilon_{\mathrm{re}}$ corresponds, in the framework of the WMAP 5yr data, to a typical redshift $z_{\mathrm{re}}= 11 \pm 1.4$.  In the $\Lambda$CDM paradigm
the temperature and polarization inhomogeneities are solely sourced by  
the fluctuations of the spatial curvature ${\mathcal R}$ whose Fourier modes obey
\begin{equation}
\langle {\mathcal R}(\vec{k}) {\mathcal R}(\vec{p}) \rangle = \frac{2\pi^2}{k^3} \delta^{(3)}(\vec{k} + \vec{p}), \qquad 
{\mathcal P}_{{\mathcal R}}(k)  = {\mathcal A}_{\mathcal R} \biggl(\frac{k}{k_{\mathrm{p}}}\biggr)^{n_{\mathrm{s}}-1},
\label{int1a}
\end{equation}
where ${\mathcal A}_{{\mathcal R}} = (2.41 \pm 0.11) \times 10^{-9}$ is the amplitude 
of the curvature power spectrum at the pivot scale  $k_{\mathrm{p}} = 0.002\,\mathrm{Mpc}^{-1}$. The success of the $\Lambda$CDM paradigm suggests the possibility of constraining also other (non-standard) parameters such as a background of relic gravitons \cite{WMAP5}, non-adiabatic modes possibly present 
prior to equality \cite{HJ} or even a background of relic magnetic fields whose presumed existence is not in contrast with the actual observations of large-scale magnetism  in galaxies \cite{gal}, clusters \cite{cl} or even (with mandatory caveats) in superclusters \cite{scl}. Can we falsify, at a given confidence level, the hypothesis that large-scale magnetic fields are already present prior to decoupling? 

This paper follows a series of investigations \cite{max1} where large-scale magnetic fields have been 
consistently included at all the stages of the Boltzmann hierarchy and within a faithful plasma description incorporating the evolution of the space-time curvature and of its inhomogeneities.  For frequencies smaller than the (electron) plasma frequency,
 the Ohmic current is solenoidal and the electric fields are suppressed by (Coulomb) conductivity: in this regime the effective dynamical variable is the centre of mass velocity of the electron ion fluid (sometimes named, with confusing terminology, the baryon velocity).  For frequencies larger  than the plasma frequency,  the propagation of
electromagnetic disturbances follows a two-fluid approach where ions and electrons are treated as independent dynamical entities: Faraday effect 
induces, in this regime,  a B-mode polarization \cite{max2} which might be relevant when (and if) compelling experimental determinations of the B-mode autocorrelation will be available.
 
The temperature autocorrelations (TT spectra in what follows) and the 
temperature-polarization cross-correlations (TE spectra in what follows) 
are determined with reasonable accuracy but,
so far, there has not been any attempt to extract the magnetic field parameters from the TT and TE angular power spectra. The modest purpose of the present investigation is to fill such a gap by using the WMAP 5yr data release in the light of the minimal version of the magnetized $\Lambda$CDM scenario (m$\Lambda$CDM in what follows) where the initial conditions of the Einstein-Boltzmann hierarchy are assigned in terms of a magnetized adiabatic mode \cite{max1} and the (magnetized) TT and TE correlations are determined accordingly. The characteristic parameters of the (minimal) m$\Lambda$CDM scenario  are the magnetic spectral index $n_{\mathrm{B}}$ and the (comoving) magnetic field intensity $B_{\mathrm{L}}$ whose associated energy density can be referred to the photon background  
as $\overline{\Omega}_{\mathrm{B L}} = B_{\mathrm{L}}^2/(8\pi \overline{\rho}_{\gamma})$. Dividing $\overline{\Omega}_{\mathrm{B L}}$ by ${\mathcal A}_{{\mathcal R}}$ we get
\begin{equation} 
\frac{\overline{\Omega}_{\mathrm{B L}}}{{\mathcal A}_{\mathcal R}}=  39.56\, \biggl(\frac{B_{\mathrm{L}}}{\mathrm{nG}}\biggr)^{2} \, \biggl(\frac{T_{\gamma0}}{2.725\,\mathrm{K}}\biggr)^{-4} \, \biggl(\frac{{\mathcal A}_{{\mathcal R}}}{2.41\times 10^{-9}}\biggr)^{-1},
\label{PS6}
\end{equation}
where $T_{\gamma 0} = 2.725 \pm 0.001\, \mathrm{K}$ is the temperature of the CMB and the comoving magnetic field $B_{\mathrm{L}}$ is measured in nG (i.e. $1\,\mathrm{nG} = 10^{-9}\,\mathrm{Gauss} \equiv 6.924 \times10^{-29}\, \mathrm{GeV}^2$). 
Since the magnetic fields are stochastically distributed,  the ensemble average of their Fourier modes obeys:
\begin{equation}
\langle B_{i}(\vec{k}) \, B_{j}(\vec{p}) \rangle = \frac{2\pi^2 }{k^3} P_{ij}(k) {\mathcal P}_{\mathrm{B}}(k) \delta^{(3)}(\vec{k} + \vec{p}), \qquad {\mathcal P}_{{\mathrm{B}}}(k) = {\mathcal A}_{\mathrm{B}} \biggl(\frac{k}{k_{\mathrm{L}}}\biggr)^{n_{\mathrm{B}}-1}, 
\label{PS2}
\end{equation}
where $P_{ij}(k) = (k^2 \delta_{ij} - k_{i} k_{j})/k^2$; 
${\mathcal A}_{\mathrm{B}}$ the spectral amplitude of the magnetic field at the pivot scale $k_{\mathrm{L}} = \mathrm{Mpc}^{-1}$ \cite{max1,max2}.  In the case when  $n_{\mathrm{B}} > 1$ (i.e. blue magnetic field spectra),  ${\mathcal A}_{\mathrm{B}} = 
(2\pi)^{n_{\mathrm{B}} -1} \, B_{\mathrm{L}}^2 /\Gamma[(n_{\mathrm{B}} -1)/2]$; if $n_{\mathrm{B}} < 1$ 
(i.e. red magnetic field spectra), ${\mathcal A}_{\mathrm{B}} =[ (1 -n_{\mathrm{B}})/2] (k_{\mathrm{A}}/k_{\mathrm{L}})^{(1 - n_{\mathrm{B}})}B_{\mathrm{L}}^2$ where $k_{\mathrm{A}}$ is the infra-red cut-off of the spectrum. In the case of white spectra (i.e. $n_{\mathrm{B}} =1$) the two-point 
function is logarithmically divergent in real space and this is fully analog to what happens in Eq. (\ref{int1a}) when $n_{\mathrm{s}} =1$, i.e. the Harrison-Zeldovich (scale-invariant) spectrum. By selecting $k_{\mathrm{L}}^{-1}$ of the order of the Mpc 
scale the comoving field $B_{\mathrm{L}}$ represents the (frozen-in) magnetic field intensity at the onset 
of the gravitational collapse of the protogalaxy \cite{max1}. 

The evolution and regularization of the magnetic fields involve an ultraviolet cut-off $k_{\mathrm{D}}$ (related to the thermal diffusivity scale at last-scattering) as well as an infra-red cut-off $k_{\mathrm{A}}$ related to the (comoving)
angular diameter distance at last scattering.  The values of $k_{\mathrm{A}}$ and $k_{\mathrm{D}}$ do depend upon
 the best-fit parameters of Eqs. (\ref{int1})--(\ref{int1a}). 
The redshift of last scattering $z_{*}$, for typical values 
of the parameters close to the values of Eqs. (\ref{int1})--(\ref{int1a}), turns out to be:
\begin{equation}
z_{*} = 1048[ 1 + f_{1} \omega_{\mathrm{b}}^{- f_{2}}] [ 1 + g_{1} \omega_{\mathrm{M}}^{\,\,g_2}],
\qquad g_{1} = \frac{0.0783\,\omega_{\mathrm{b}}^{-0.238}}{[1 + 39.5 (\omega_{\mathrm{b}})^{0.763}]},\qquad g_{2} = \frac{0.560}{1 + 21.1 \omega_{\mathrm{b}}^{1.81}},
\label{PS4b}
\end{equation}
where $\omega_{\mathrm{M}} = \omega_{\mathrm{b}} + \omega_{\mathrm{c}}$ and $\omega_{\mathrm{b}} = 
 h_{0}^2 \Omega_{\mathrm{b}0}$, $\omega_{\mathrm{c}} = h_{0}^2 \Omega_{\mathrm{c}0}$; in Eq. (\ref{PS4b}) 
$f_{1} = 1.24\times 10^{-3}$ and $f_{2} = 0.738$. In the case of the parameters of Eqs. (\ref{int1})--(\ref{int1a}), 
Eq. (\ref{PS4b}) implies $z_{*} =1090.5$ in excellent agreement with the estimate of the WMAP collaboration \cite{WMAP5}, i.e. $z_{*} = 1090.51 \pm 0.95$. From the value of $z_{*}$ it is possible 
to compute $r_{\mathrm{R}*}= \rho_{\mathrm{R}}(z_{*}) / \rho_{\mathrm{M}}(z_{*}) = 4.15 \times 10^{-2}/\omega_{\mathrm{M}} 
\, (z_{*}/10^{3})$.  Recalling that  the comoving angular diameter distance can be written as 
$D_{A}(z_{*}) = 2 d_{\mathrm{A}}(z_{*})/(H_{0} \sqrt{\Omega_{\mathrm{M}0}})$, $k_{\mathrm{A}}(z_{*})$ and 
$k_{\mathrm{D}}(z_{*})$ are determined in terms of Eqs. (\ref{int1})--(\ref{int1a}) 
\begin{equation}
k_{\mathrm{A}}(z_{*}) = 1/D_{\mathrm{A}}(z_{*}),\qquad \frac{k_{\mathrm{D}}(z_{*})}{k_{\mathrm{A}}(z_{*})}= \frac{2240 \, d_{\mathrm{A}}(z_{*})}{\sqrt{\sqrt{r_{\mathrm{R}*} +1} - \sqrt{r_{\mathrm{R}*}}}} 
\biggl(\frac{z_{*}}{10^{3}} \biggr)^{5/4} \, \omega_{\mathrm{b}}^{0.24} \omega_{\mathrm{M}}^{-0.11},
\label{PS4a}
\end{equation}
where, for the best fit parameters listed in Eqs. (\ref{int1})--(\ref{int1a}), $d_{\mathrm{A}}(z_{*}) =0.8569$ leading to 
$D_{\mathrm{A}}(z_{*}) = 14110.8 \, \mathrm{Mpc}$ in excellent agreement with the WMAP 5yr determination
(i.e. $D_{\mathrm{A}}(z_{*}) = 14115_{-191}^{188}\, \mathrm{Mpc}$) implying
 $k_{\mathrm{A}}(z_{*}) = 7.08\times 10^{-5} \, \mathrm{Mpc}^{-1}$. 

Throughout the numerical scan of the parameter space the initial conditions of the Boltzmann hierarchy have been given in terms of the magnetized adiabatic mode which has been discussed in \cite{max1,max2}. The numerical  code employed here is an improved version of the one used in \cite{max1,max2} and it is based on Cmbfast \cite{cmbf} (which follows, in turn, Cosmics \cite{cosm}).  Let us now remark that the joined two-dimensional marginalized contours for the various cosmological parameters identified already by the analyses of the WMAP 3yr data  \cite{WMAP3} 
are ellispses with an approximate Gaussian dependence on the confidence level as they must be in the Gaussian approximation (see, e.g., already the first paper of Ref. \cite{WMAP3} and, in particular, Fig. 10).
In the light of this observation it is then plausible to determine the confidence 
intervals for the $2$ supplementary  parameters of the model (i.e. $n_{\mathrm{B}},\, B_{\mathrm{L}}$) by using an appropriate gridding  approach while the remaining parameters of Eqs. (\ref{int1})--(\ref{int1a}) are assumed to be known and follow a Gaussian probability density function. This approach has been also followed when discussing, some time ago, the impact of non-adiabatic 
modes on the WMAP 3yr data (see, e. g.,  second paper of Ref. \cite{HJ}). In the numerical fit to the data the shape of the likelihood function can be determined by evaluating the least square estimator
\begin{equation}
\chi^2(n_{\mathrm{B}}, B_{\mathrm{L}}) = 
\sum_{\ell} \biggl[\frac{ C_{\ell}^{(\mathrm{obs})} - C_{\ell}(n_{\mathrm{B}},  B_{\mathrm{L}})}{\sigma_{\ell}^{(\mathrm{obs})}} \biggr]^2, 
\label{grid1}
\end{equation}
where $\sigma_{\ell}^{\mathrm{obs}}$ are the estimated errors from the observations 
for each multipole and where the functional dependence of  $C_{\ell}(n_{\mathrm{B}}, B_{\mathrm{L}})$ is given by the underlying theory (i.e. the magnetized $\Lambda$CDM model) which we try to falsify by comparing its predictions to the actual observations.  The observed angular power spectra 
(i.e. $C_{\ell}^{\mathrm{obs}}$) are derived by using the (further) estimators $\hat{C}_{\ell}^{(\mathrm{TT})}$ and $\hat{C}_{\ell}^{(\mathrm{TE})}$, i.e.  
\begin{equation}
\hat{C}_{\ell}^{(\mathrm{TT})}= \frac{1}{2 \ell +1} \sum_{m = - \ell}^{\ell} |\hat{a}^{(\mathrm{T})}_{\ell \, m}|^2, \qquad 
 \hat{C}_{\ell}^{(\mathrm{TE})}= \frac{1}{2 \ell +1} \sum_{m = - \ell}^{\ell} |\hat{a}^{(\mathrm{T})}_{\ell \, m}\,\hat{a}^{(\mathrm{E})*}_{\ell \, m}|, 
\label{grid2}
\end{equation}
whose distribution becomes Gaussian, according to the central limit theorem, when $\ell \to \infty$.  
The true  $a^{(\mathrm{T})}_{\ell\, m}$ and $a^{(\mathrm{E})}$ are 
not accessible to direct observation, however, their estimates (i.e. $\hat{a}^{(\mathrm{T})}_{\ell\, m}$ and 
$\hat{a}^{(\mathrm{E})}_{\ell\, m}$) can be directly inferred from the measured temperature and polarization inhomogeneities \cite{WMAP5}. 
The minimization of Eq. (\ref{grid1}) is equivalent to the minimization of the lognormal likelihood function ${\mathcal L} = - 2 \, \ln{L}$ where $L$ is given by
\begin{equation}
L(\mathrm{data}|\,n_{\mathrm{B}}, \, B_{\mathrm{L}}) = L_{\mathrm{max}}
e^{- \chi^2(n_{\mathrm{B}}, B_{\mathrm{L}})/2}.
\label{grid4}
\end{equation}
Thus, the minimization of Eq. (\ref{grid1})  is equivalent to the maximization of the likelihood of Eq. (\ref{grid4}). 
\begin{figure}[!ht]
\centering
\includegraphics[height=7cm]{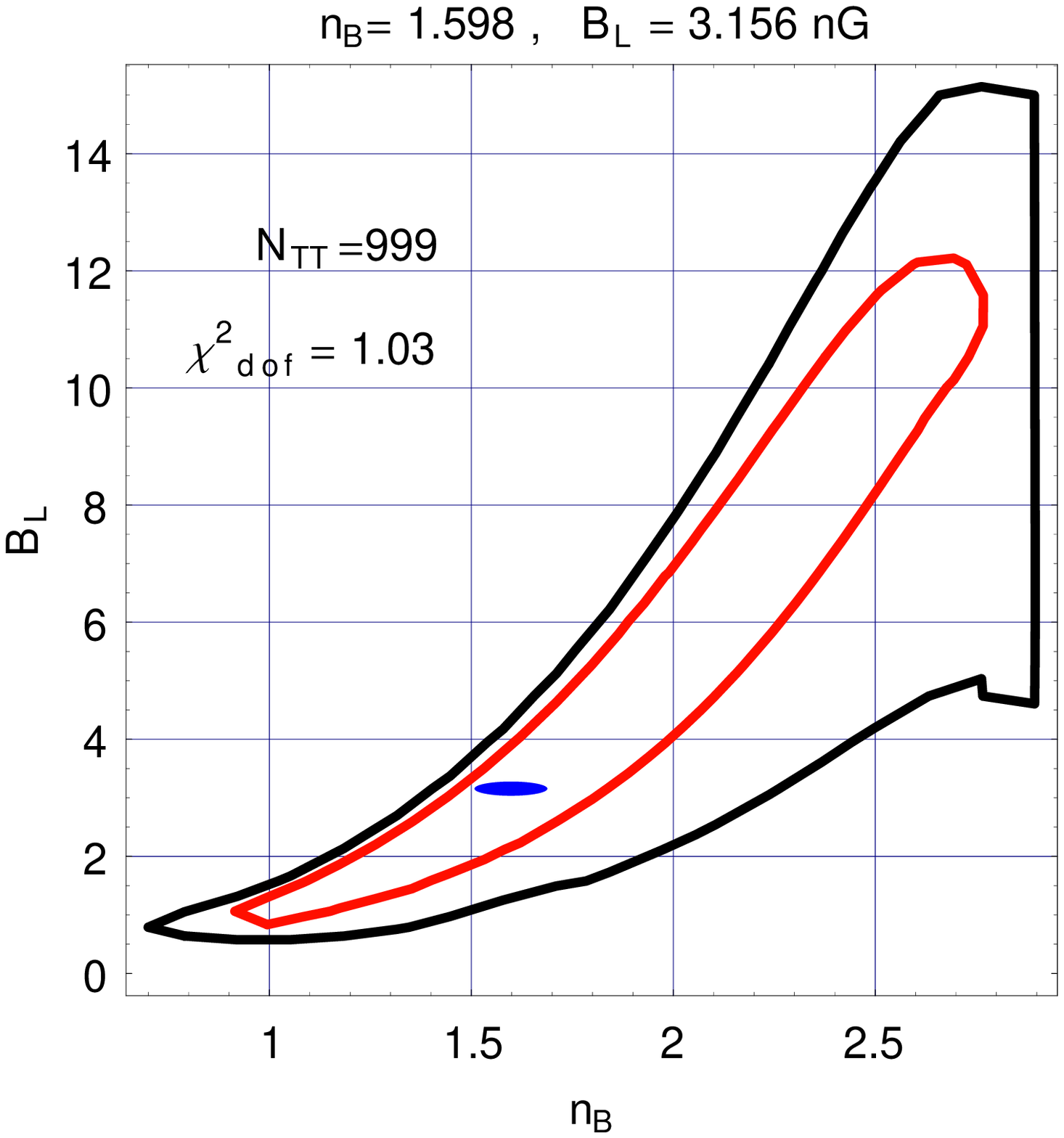}
\includegraphics[height=7cm]{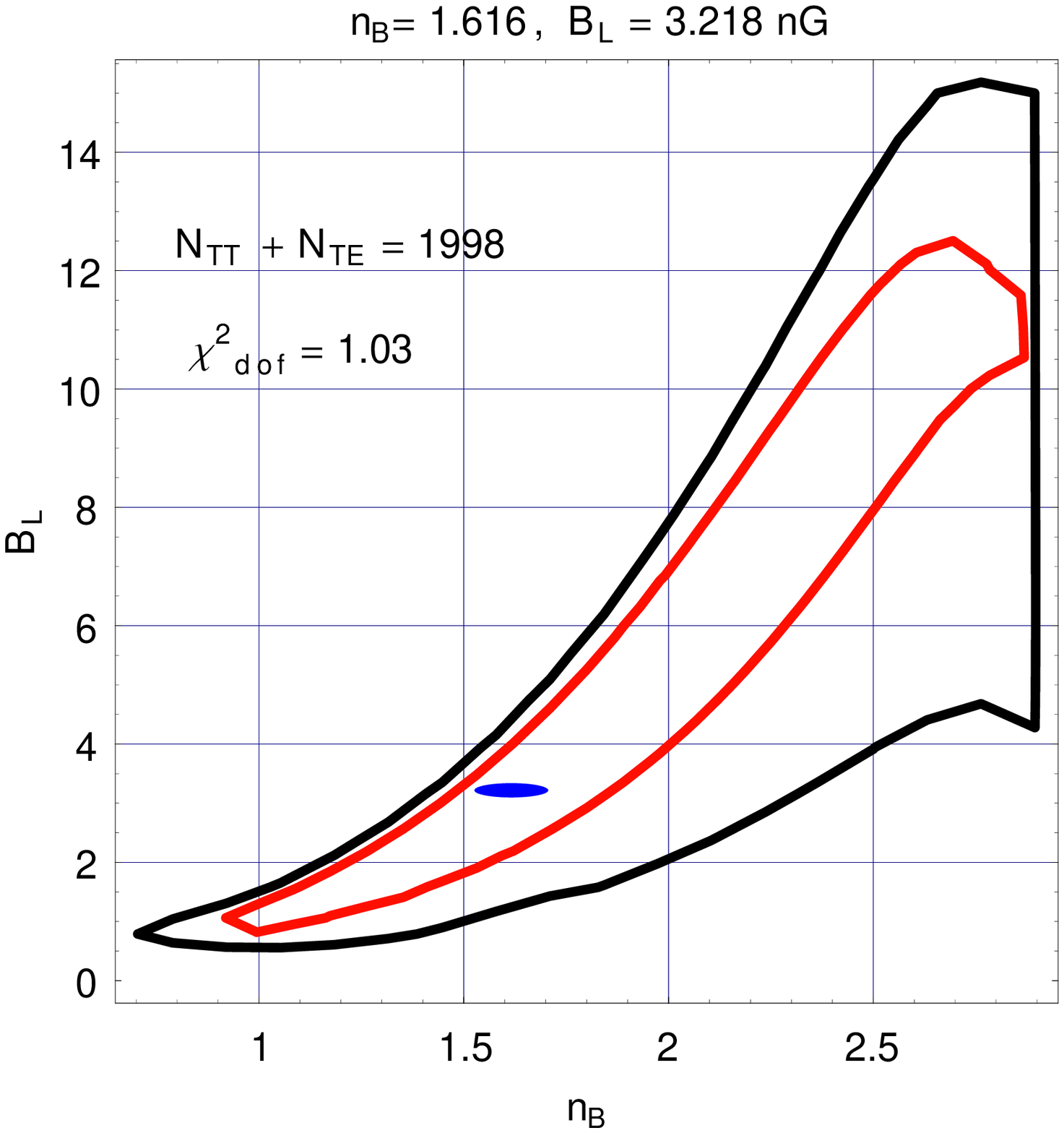}
\caption[a]{Likelihood contours in the plane $(n_{\mathrm{B}},\, B_{\mathrm{L}})$. In both plots the inner curve corresponds to $\Delta \chi^2 = 2.3$ (i.e. $68.3$\% or $1\, \sigma$ of likelihood content) while the outer curve  corresponds to $\Delta \chi^2 = 6.17$ (i.e. $95.4\,$\% or $2\, \sigma$ of likelihood content).}
\label{FIG1}      
\end{figure}
In Fig. \ref{FIG1} the 
filled (ellipsoidal) spots in both plots highlight the minimal value of the $\chi^2$ (i.e. 
$\chi^2_{\mathrm{min}}$) while the contour plots represent the
likelihood contours in the two parameters $n_{\mathrm{B}}$ and $B_{\mathrm{L}}$.
In both plots of Fig. \ref{FIG1} the boundaries of the two regions 
contain $68.3\,$\% and $95.4\,$\% of likelihood as the 
values for which the $\chi^2$ has increased, respectively, by an amount $\Delta \chi^2 = 2.3$ and $\Delta \chi^2 = 6.17$. In the plot at the left of Fig. \ref{FIG1} the data points correspond to the measured TT correlations contemplating 
$N_{\mathrm{TT}} =999$ experimental points from $\ell = 2$ to $\ell =1000$.
 In the plot at the right the data points include, both, the TT and TE correlations 
 and the total number of data points increases to $N_{\mathrm{TT}} + N_{\mathrm{TE}} =1998$.
The values of $n_{\mathrm{B}}$ and $B_{\mathrm{L}}$ minimizing the $\chi^2$ when 
only the TT correlations are considered  turn out to be $n_{\mathrm{B}}= 1.598$ and $B_{\mathrm{L}}= 3.156\, \mathrm{nG}$ (see 
Fig. \ref{FIG1} plot at the left); in this case the 
reduced $\chi^2$ is $1.09$. When also the TE correlations are included in the 
analysis the reduced $\chi^2$ diminishes from $1.09$ to $1.03$ and the 
values of $n_{\mathrm{B}}$ and $B_{\mathrm{L}}$ minimizing the $\chi^2$
become $n_{\mathrm{B}} =1.616$ and  $B_{\mathrm{L}} =3.218\, \mathrm{nG}$. In the frequentist approach, the boundaries of the confidence regions represent exclusion plots at  $68.3\,$\% and $95.4\,$\% confidence level.  In this sense the regions beyond
the outer curves in Fig. \ref{FIG1} are excluded to $95$\% confidence level.

\begin{figure}[!ht]
\centering
\includegraphics[height=6cm]{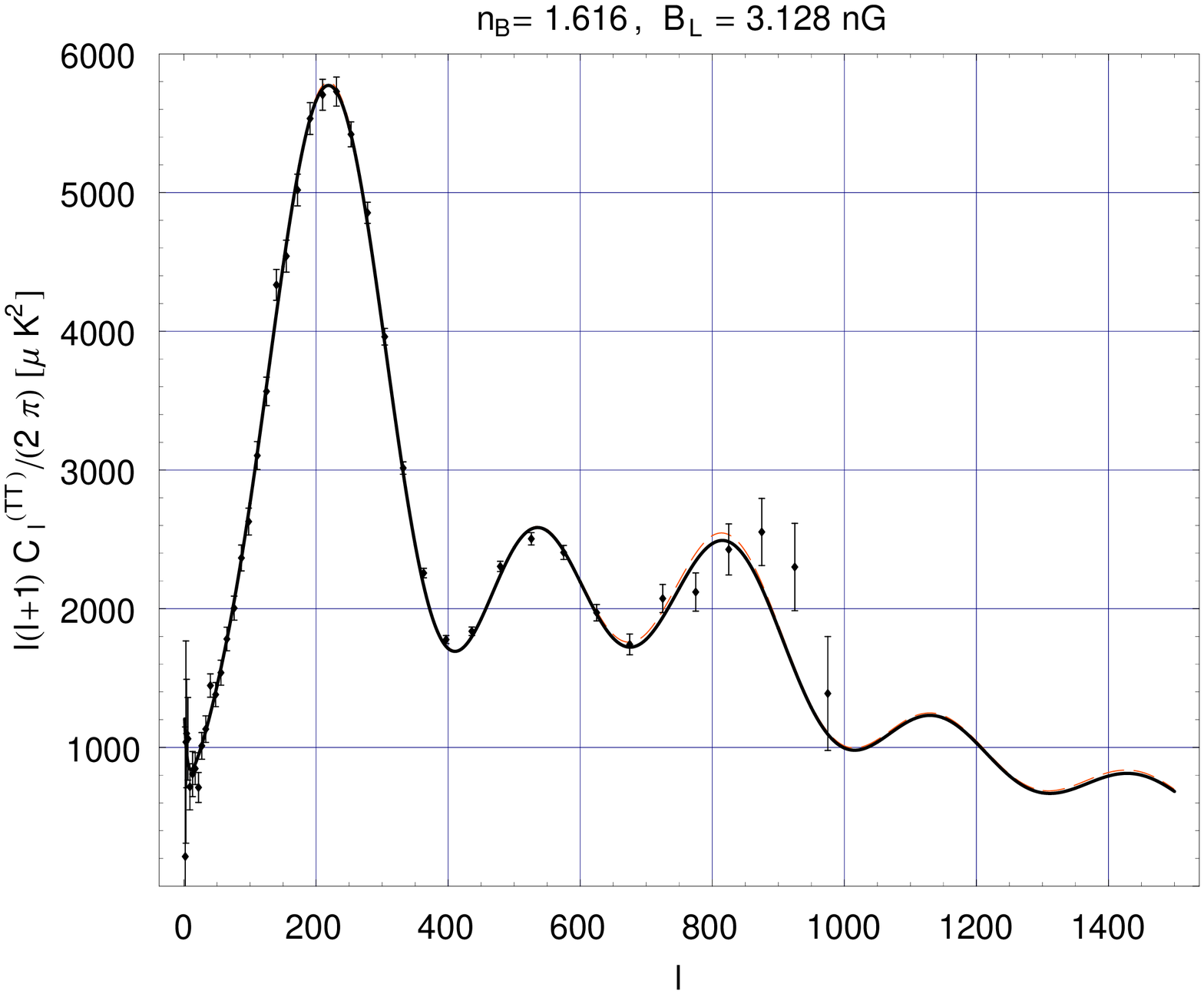}
\includegraphics[height=6cm]{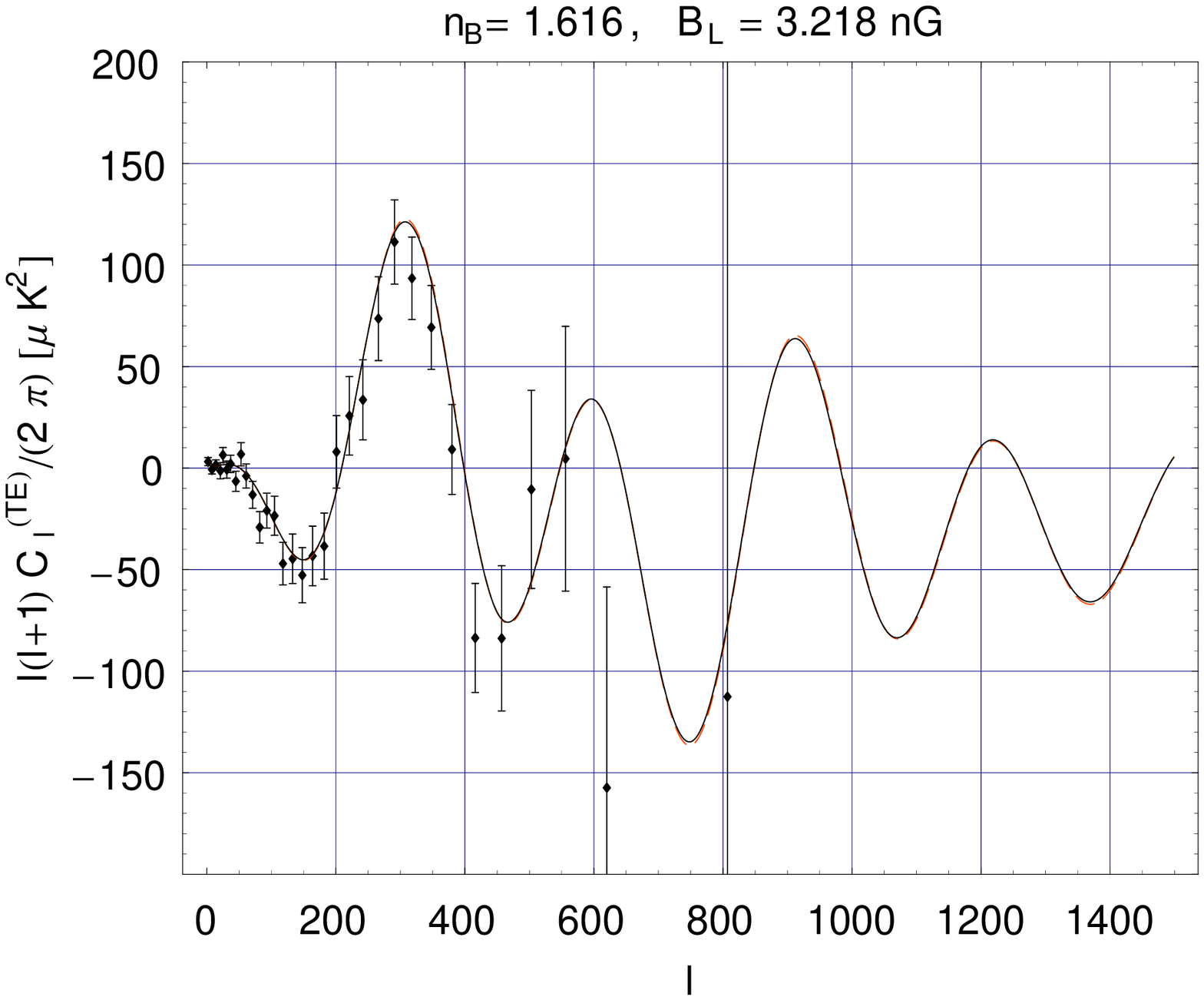}
\caption[a]{The TT and TE angular power 
spectra for the best-fit parameters 
stemming from the right plot of Fig. \ref{FIG1} (dashed line) and from the best 
fit to the WMAP 5yr data alone (full line). The data points of the WMAP 5yr release are included in their binned form.}
\label{FIG2}      
\end{figure}

\begin{figure}[!ht]
\centering
\includegraphics[height=6.5cm]{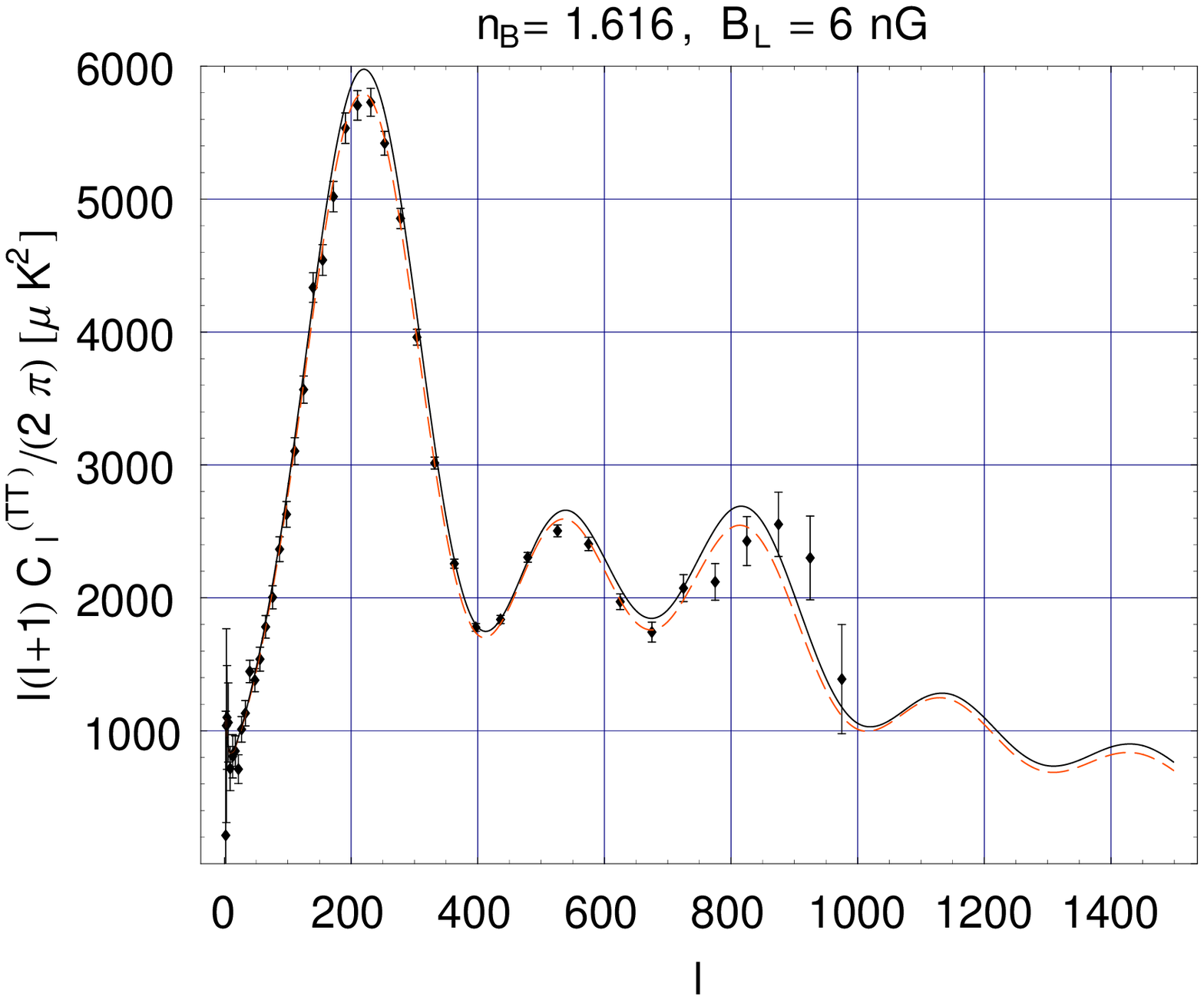}
\includegraphics[height=6.5cm]{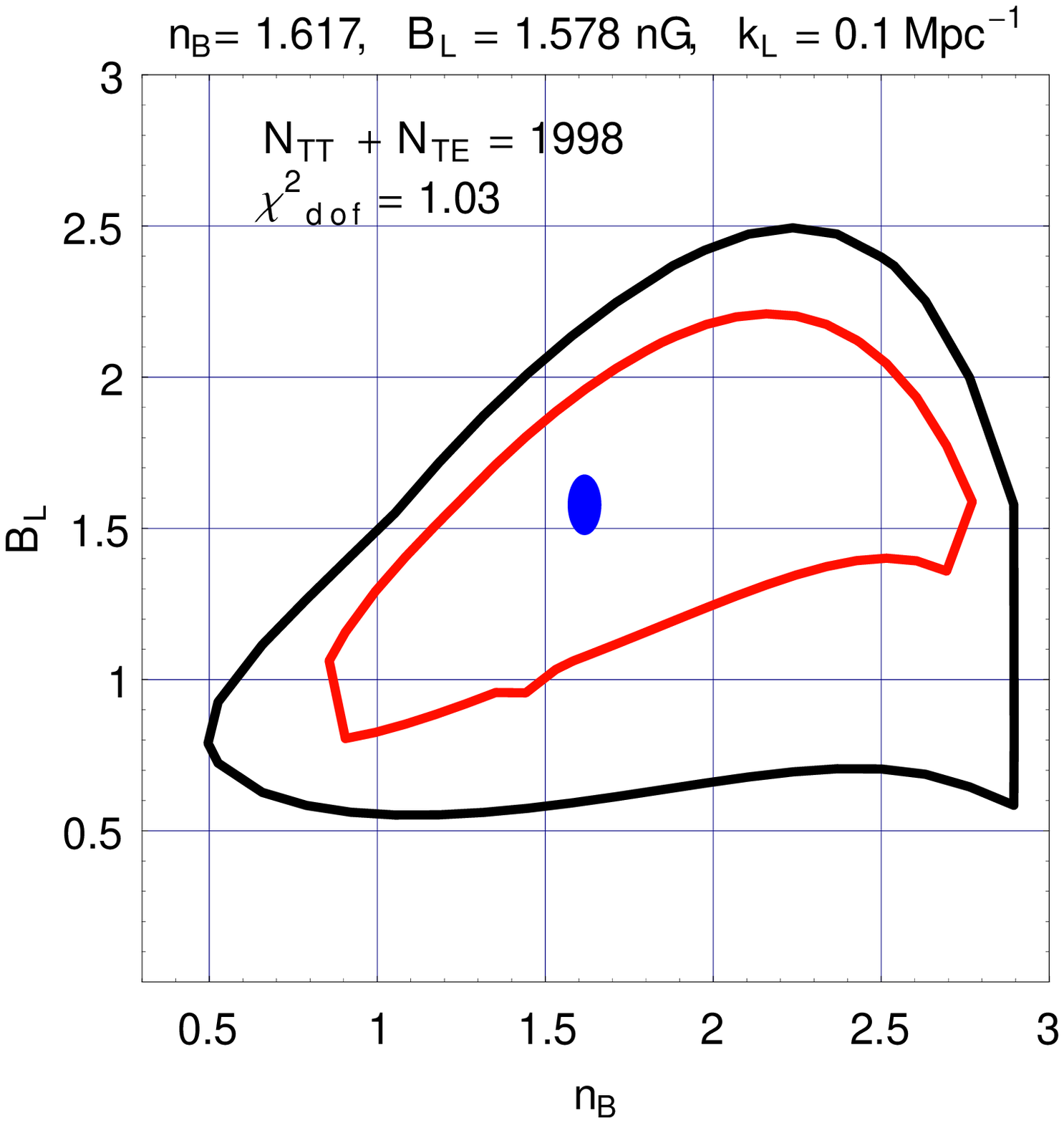}
\caption[a]{Left plot: the TT power spectra which are more than $2 \sigma$ away from the best-fit point (full line) are compared with the best fit of Fig. \ref{FIG2} (dashed line). Right plot: contour plots in the $(n_{\mathrm{B}}, B_{\mathrm{L}})$ plane 
for $k_{\mathrm{L}} = 0.1 \, \mathrm{Mpc}^{-1}$ (see also Fig. \ref{FIG1} where $k_{\mathrm{L}} = 1\, \mathrm{Mpc}^{-1}$).}
\label{FIG3}      
\end{figure}

In Fig. \ref{FIG2} the full line corresponds to the WMAP 5yr best fit of Eqs. (\ref{int1})--(\ref{int1a}).
The (hardly distinguishable) dashed lines, in both plots, illustrate the 
m$\Lambda$CDM result for $(n_{\mathrm{B}}=1.616,\, B_{\mathrm{L}}=3.218\, \mathrm{nG})$ corresponding to $\chi^2_{\mathrm{min}}$, as 
determined from the joined  analysis of the TT and TE angular power spectra (see plot at the right in Fig. \ref{FIG1}). In both plots of Fig. \ref{FIG1} 
the (binned) data points are reported \footnote{
The unbinned data (which are the ones used in the analysis) contemplate all the 
 multipoles from $\ell =2$ to $\ell = 1000$ both for the TT and for the TE (observed)
 power spectra. The binned data contain 
instead $34$ (effective) multipoles in the TE correlation and $43$ (effective)multipoles in the TT spectrum. Following the usual habit, to make the plots more readable,  the binned data points have been included in Fig. \ref{FIG2}.}.
The impact of the magnetic field parameters can be appreciated by monitoring
 the relative ratios of the first three peaks. Using the notation ${\mathcal G}_{\ell} =
 \ell (\ell +1) C_{\ell}/(2\pi)$, the relative heights of the acoustic peaks computed 
 in the case of the best-fit of Fig. \ref{FIG2} are:
\begin{equation}
\overline{H}_{1} = \frac{{\mathcal G}_{\ell_{1}}}{{\mathcal G}_{\ell =10}} = 6.942,\qquad \overline{H}_{2} = \frac{{\mathcal G}_{\ell_{2}}}{{\mathcal G}_{\ell_{1}}} = 0.447, \qquad 
\overline{H}_{3} = \frac{{\mathcal G}_{\ell_{3}}}{{\mathcal G}_{\ell_{2}}} = 0.981,
\label{PS8}
\end{equation}
where $\ell_{1} = 220$,  $\ell_{2} = 535$ and  $\ell_{3} = 816$ are, respectively, the locations 
of the first three acoustic peaks. In the the case of Eqs. (\ref{int1})--(\ref{int1a}) (i.e. in the absence of a magnetized 
background) the same ratios are, respectively, 
$\overline{H}_{1}= 6.876$, $\overline{H}_{2} = 0.447$ and $\overline{H}_{3} = 0.963$.
In Fig. \ref{FIG3} (plot at the left) the TT power spectra are illustrated 
with the same value of the spectral index as in Fig. \ref{FIG2} (i.e. $n_{\mathrm{B}} = 1.616$) 
but with a magnetic field intensity $B_{\mathrm{L}} = 6$ nG which is more 
than two $\sigma$ away from $\chi^2_{\mathrm{min}}$ (see Fig. \ref{FIG1}, plot at the right): as 
expected,  rough inspection of the left plot in Fig. \ref{FIG3}, reveals that the  m$\Lambda$CDM estimate does not correctly reproduce the data when ($n_{\mathrm{B}} = 1.616$, $B_{\mathrm{L}} = 6$ nG) especially around the first and third peaks. In the latter case the the relative ratios of the peaks  become
  $(\overline{H}_{1}, \overline{H}_{2}, \overline{H}_{3}) = ( 7.142, 0.444, 1.012)$: i.e. 
  $\overline{H}_{1}$ and $\overline{H}_{3}$  exceed the best fit while $\overline{H}_{2}$ is comparatively lower than in Eq. (\ref{PS8}). This is a general trend in the excluded region.
Consequently, the whole approach of \cite{max1} is rather effective in constraining 
variations of $B_{\mathrm{L}}$ in the nG range. Similarly, by moving the magnetic 
spectral index in the excluded region of the parameter space the distortions of the peaks jeopardize the goodness 
of the fit.
In Fig. \ref{FIG3} (plot at the right) the magnetic pivot wave-number  has been 
moved from $k_{\mathrm{L}} = 1\, \mathrm{Mpc}^{-1}$ to 
$k_{\mathrm{L}} = 0.1\, \mathrm{Mpc}^{-1}$, consistently with the indetermination of the scale 
of protogalactic collapse. A reduction of the pivot 
wave-number implies that the magnetic field is regularized 
over a broader window in real space; hence the allowed amplitude 
of the putative magnetic field intensity is reduced. In the right plot of Fig. \ref{FIG3} 
the $\chi^2$ is minimized when $n_{\mathrm{B}} = 1.617$ (i.e. comparable with the value of $n_{\mathrm{B}}$ deduced in Fig. \ref{FIG2} when $k_{\mathrm{L}} = 1 \,\mathrm{Mpc}^{-1}$) and, as anticipated, for a smaller magnetic field, i.e. $B_{\mathrm{L}}= 1.578 \, \mathrm{nG}$.  The modification in the 
pivot wave-number also modifies the shape of the confidence contours (compare, e.g., Fig. \ref{FIG1} and Fig. \ref{FIG3}).  

The large-scale magnetic field possibly present 
prior to decoupling have been estimated, for the first time, from the temperature autocorrelations 
as well as from the cross-correlations between temperature and polarization in the case 
of the magnetized adiabatic mode \cite{max1}. The obtained results prove
in explicit terms a guessed degeneracy between the spectral 
index and the magnetic field intensity. In a frequentistic 
perspective, beyond the outer contours of Figs. \ref{FIG1} and \ref{FIG3} 
the parameter space of the (minimal) m$\Lambda$CDM scenario is excluded to $95$\% confidence level.
Global observables (such as, for instance, the heights 
and depths of the acoustic oscillations) could be probably 
used as a normal parameter basis to infer an explicit analytic dependence of the TT and TE correlations upon the parameters 
of the magnetized background. In this direction work is in progress.

\end{document}